\documentclass[12pt,notitlepage,fleqn]{article}
\usepackage{graphicx}
\usepackage{amsmath}
\usepackage{indentfirst}

\newtheorem{problem}{Hypothesis}

\begin{document}

\begin{center}
{\LARGE A Phenomenological Model of the Baryons}

The Body Center Cubic Model of the Vacuum Material

\bigskip

{\normalsize Jiao Lin Xu}

{\small The Center for Simulational Physics, The Department of Physics and
Astronomy}

{\small University of Georgia, Athens, GA 30602, U. S. A.}

E- mail: {\small \ Jxu@Hal.Physast.Uga.edu}

\bigskip

\textbf{Abstract}
\end{center}

{\small From the quark confinement idea, we conjecture that the quarks
compose colorless particles (uud and udd - the} \textbf{Lee Particles}%
{\small ) and then the Lee Particles construct\ a body center cubic lattice
in the vacuum. In terms of the energy band theory, from the symmetries of
the body center cubic periodic field, we deduce the baryon spectrum (with 
\textbf{a united mass formula}) using only \textbf{2 flavored quarks} u and
d. We also predict some new baryons: }$\Lambda ^{0}(2559),${\small \ }$%
\Lambda _{C}^{+}(6659)${\small , }$\Lambda _{b}^{0}(10159)${\small .... The
experiments to find the long lifetime baryon }$\Lambda ^{0}(2559)${\small \
should be done first.}

\section{Introduction}

The Quark Model \cite{QuarkModel} has already explained the baryon spectrum
in terms of quarks. It successfully gives intrinsic quantum numbers ($I$, $S$%
, $C$, $b$, and $Q$) of all baryons. However, (1) it has not given a
satisfactory mass spectrum of baryons in a united mass formula \cite
{RELATION of MASS}; (2) it needs too many elementary particles (6 flavors $%
\times $ 3 colors $\times $ 2 (quark and antiquark) = 36 quarks) \cite
{QUARKS} \cite{Three Color}; (3) the quantum numbers of the quarks are
``entered by hand'' \cite{Vacuum engineering} \cite{Hand in}; (4) on one
hand it assumes \cite{QuarkModel} that all quarks (u, d, s, c, b, t) are
independent elementary particles, but on the other hand it assumes that the
higher energy quarks can decay into lower energy quarks \cite{QUARK DECAY},
the two ``hands'' do not cooperate with each other; (5) all free quark
searches since 1977 have had negative results \cite{Free QUARK}. Just as T.
D. Lee pointed out \cite{Marshah}: ``In order to apply the present theories,
we need about seventeen ad hoc parameters. All these theories are based on
symmetry considerations, yet most of the symmetry quantum numbers do not
appear to be conserved. All hadrons are made of quarks and yet no single
quark can be individually observed. Now,fifty years after the beginning of
modern particle physics, our successes have brought us to the deeper
problems. We are in a serious dilemma about how to make the next giant step.
Because the challenge is related to the very foundation of the totality of
physics, a breakthrough is bound to bring us a profound change in basic
science.'' This paper trys to find a solution to the above problems.

Twenty years ago, T. D. Lee had already pointed out \cite{Vacuum engineering}
: ``we believe our vacuum, though Lorentz invariant, to be quite
complicated. Like any other physical medium, it can carry long-range-order
parameters and it may also undergo phase transitions... .'' Recently, Frank
Wilczek, the J. Robert Oppenheimer Professor at the Institute for Advanced
Study in Princeton, further elaborated Lee's idea \cite{wilczek}: ``empty
space--the vacuum--is in reality a richly structured, though highly
symmetrical, medium. Dirac's sea was an early indication of this feature,
which is deeply embedded in quantum field theory and the Standard Model.
Because the vacuum is a complicated material governed by locality and
symmetry, one can learn how to analyze it by studying other such
materials--that is, condensed matter.'' Professor Wilczek not only pointed
out one of the most important and most urgent research directions of modern
physics--studying the structure of the vacuum, but also provided a very
practical and efficient way for the study--learning from studying condensed
matter. Applying the Lee-Wilczek idea, this paper conjectures a structure of
the vacuum (body center cubic symmetry), which will be used as the mechanism
to generate the baryon\ spectrum \cite{NetXu}.

According to Dirac's sea concept \cite{diracsea}, there are follow Dirac
seas: electron sea, $\mu $ lepton sea, $\tau $ lepton sea, $u$ quark sea, $d$
quark sea, $s$ quark sea, $c$ quark sea, $b$ quark sea...in the vacuum. All
of these Dirac seas are in the same space, at any location, that is, at any
physical space point. \textbf{The facts (all hadrons are made of quarks and
no single quark can be individually observed) imply that the quarks are
confined in hadrons in the vacuum.} According to quantum chromodynamics \cite
{Chromodynamics}, there are\textbf{\ super-strong color attractive
interactions} among the quarks, causing three quarks of different colors to
be confined together and form a colorless baryon ($p$, $n$, $\Lambda $, $%
\Sigma $, $\Xi $, $\Omega ..$) in the vacuum. These baryons, electrons,
leptons, etc. will interact with one another and form the perfect vacuum
material. However, some kinds of particles do not play an important role in
forming the vacuum material. First, the main force which makes and holds the
structure of the vacuum material must be the strong interactions, not the
weak-electromagnetic, or the gravitational interactions. Hence, in
considering the structure of the vacuum material, we leave out the Dirac
seas of those particles which do not have strong interactions ($e,$ $\mu $, $%
\tau $). Secondly, it is unlikely that the super stable vacuum material is
composed of unstable blocks (the unstable baryons are short lived), hence we
also omit the unstable particles (such as: $\Lambda $, $\Sigma $, $\Xi $, $%
\Omega $, ...). Finally, there are only two kinds of possible particles
left: the vacuum state protons (uud-\textbf{charged Lee Particle }\cite
{Vacuum engineering} \cite{Marshah} \cite{asymmetry} \cite{li pARTICLE}) and
the vacuum state neutrons (udd-\textbf{neutral Lee Particle}). It is well
known that there are strong attractive forces between the protons and the
neutrons inside a nucleus. Similarly, there should also exist strong
attractive forces between the Lee Particles which will make and hold the
densest structure of the vacuum state Lee Particles.

According to solid state physics \cite{Solidstates}, if two kinds of
particles (with radius $R_{1}<R_{2}$) satisfy the condition $%
1>R_{1}/R_{2}>0.73$, the densest structure is the body center cubic crystal 
\cite{bodycenter}. According to the Quark Model, the charged Lee Particle
(uud) and the neutral Lee Particle (udd) are not completely the same, thus $%
R_{1}\neq R_{2}$; and they are similar to each other, thus $R_{1}\approx
R_{2}$. Hence, if $R_{1}<R_{2}$ (or $R_{2}<R_{1}$), we have $%
1>R_{1}/R_{2}>0.73$ (or $1>R_{2}/R_{1}>0.73$). Therefore, we conjecture that
the vacuum state Lee Particles construct the densest structure-\textbf{a
body center cubic lattice }(in this paper it will be regarded as \textbf{the
BCC model}) in the vacuum.

Similar to a crystal which has a periodic field, there are also periodic
fields in the vacuum. From energy band theory \cite{eband} and the
phenomenological fundamental hypotheses of the BCC model, we can deduce all
intrinsic quantum numbers of all baryons which are consistent with the
experimental results \cite{particle}. Likewise, we can calculate the masses
of all baryons which are in very good agreement with the experimental
results \cite{particle} using \textbf{a united mass formula}.

The vacuum material works like an \textbf{ultra-superconductor}.\ Since the
energy gaps are so large (for electron the energy gap is about 0.5 Mev; for
proton and neutron the energy gaps are about 939 Mev), there is no electric
and mechanical resistance to any particle and any physical body (made by
protons, neutrons, and electrons) moving inside the vacuum material with
constant velocity.

\section{Fundamental Hypotheses}

For simplicity, the intrinsic structure (three quarks) of baryons will be
ignored temporarily. The baryons are treated as elementary particles in the
phenomenological BCC model. We would like to call this simplification 
\textbf{the} \textbf{point baryon approximation}.\textbf{\ }The
approximation is based on the the quark confinement theory \cite{CONFINEMENT}
and the experimental results \cite{Free QUARK} that a baryon always appears
as a whole particle.

In order to explain our model accurately and concisely, we will start from
the phenomenological fundamental hypotheses in an axiomatic form.

\begin{problem}
There are only two kinds of fundamental quarks u and d in the quark family.
There exist super-strong color attractive interactions between the\ colored
quarks. Three quarks (uud or udd) compose a kind of colorless Fermi particles%
\textbf{\ in the vacuum state}. We call the particles the \textbf{Lee
pareticles }\cite{Vacuum engineering} \cite{Marshah} \cite{asymmetry} \cite
{li pARTICLE}.
\end{problem}

\textbf{According to the Quark Model} \cite{QuarkModel}, the Lee Particles
are unflavored ($S=C=b=0$) with spin $s=1/2$ and isotropic spin $I=1/2.$ The
excited (from the vacuum) free Lee Particles have 
\begin{equation}
\begin{tabular}{lll}
$uud$ & B=1, S=C=b=0, s=1/2, I=1/2, I$_{z}$= 1/2 & $proton$ \\ 
$udd$ & B=1, S=C=b=0, s=1/2, I=1/2, I$_{z}$= -1/2 & $neutron$%
\end{tabular}
\label{nucleon}
\end{equation}

\begin{problem}
There are strong attractive interactions between the Lee Particles, and the
interactions will make and hold the densest structure of the Lee Particles - 
\textbf{the body center cubic Lee Particle Lattice in the vacuum.} The
lattice forms a strong interaction periodic field with body center cubic
symmetries in the vacuum, where the periodic constant $a_{x}$ is much
smaller than the magnitude of the radii of the nuclei.
\end{problem}

\begin{problem}
Quantum mechanics applies to the ultra-microscopic world \cite{TDLEE}. Thus,
the energy band theory \cite{eband} is also valid in the ultra-microscopic
world. \textbf{The energy band excited states of the Lee Particles will be
various baryons}.
\end{problem}

According to the energy band theory, an excited Lee Particle (from vacuum),
inside the body center cubic periodic field, will be in a state of the
energy bands (a point of the Brillouin zone). \textbf{The first Brillouin
zone} \cite{Brillouin} of the body center cubic lattice is shown in Fig. 1.
In Fig. 1 (depicted from \cite{eband} (Fig. 1) and \cite{Brillouin}(Fig.
8.10)), the $(\xi ,\eta ,\zeta )$ coordinates of the symmetry points are: 
\begin{gather}
\Gamma =(\text{0, 0, 0}),\text{ }H=(\text{0, 0, 1}),\text{ }P=(\text{1/2,
1/2, 1/2}),  \notag \\
N=(\text{1/2, 1/2, 0}),\text{ }M=(\text{1, 0, 0}),
\end{gather}
and the $(\xi ,\eta ,\zeta )$ coordinates of the symmetry axes are: 
\begin{eqnarray}
\Delta &=&(\text{0, 0, }\zeta ),\text{\ }0<\zeta <1;\text{ \ \ \ \ \ \ }%
\Lambda =(\xi \text{, }\xi \text{, }\xi ),\text{ }0<\xi <1/2;  \notag \\
\Sigma &=&(\xi \text{, }\xi \text{, 0}),\text{ }0<\xi <1/2;\text{ \ \ \ }D=(%
\text{1/2, 1/2, }\xi ),\text{ }0<\xi <1/2;  \notag \\
G &=&(\xi \text{, 1-}\xi \text{, 0}),\text{ }1/2<\xi <1;\text{ \ }F=(\xi 
\text{, }\xi \text{, 1-}\xi ),\text{ }0<\xi <1/2.
\end{eqnarray}
From Fig. 1, we know that the axis $\Delta (\Gamma -H)$ is a $4$-fold
rotation axis, the axis $\Lambda (\Gamma -P)$ is a $3$-fold rotation axis,
and the axis $\Sigma (\Gamma -N)$ is a $2$-fold rotation axis. \ 

\begin{problem}
Due to the effect of the periodic field, fluctuations of an excited Lee
Particle state may exist. Thus, the fluctuations of energy $\varepsilon $%
{\LARGE \ }and intrinsic quantum numbers (such as the strange number $S$)
may also exist. The fluctuation of the Strange number, if exists, is always%
\textbf{\ }$\Delta S=\pm 1$ \cite{real value of S}. From the fluctuation of
the Strange number, we will be able to deduce new quantum numbers, such as
the \textbf{Charmed number} $C$ and the \textbf{Bottom number} $b$.
\end{problem}

\begin{problem}
The energy band excited states (except for the energy bands in the first
Brillouin zone) of the Lee Particles are the unstable baryons (we call all
baryons except protons and neutrons\textbf{\ unstable baryons}). Their
quantum numbers and masses are determined as follows (note: the quantum
numbers of the ground energy bands in the first Brillouin zone are determind
by \textbf{Hypothesis I):}
\end{problem}

\begin{enumerate}
\item  Baryon number $B$: according to \textbf{Hypothesis I}, all energy
band states have 
\begin{equation}
B=1.  \label{baryon}
\end{equation}

\item  Isospin number $I$: the maximum isospin $I_{m}$ is determined by the
energy band degeneracy $d$ \cite{eband}, where 
\begin{equation}
d=2I_{m}+1,  \label{isomax}
\end{equation}

and another possible isospin value is determined by 
\begin{equation}
I=I_{m}-1,\text{ \ \ }I\geq 0.  \label{isonext}
\end{equation}
In some cases the degeneracy $d$ should be divided into sub-degeneracies
before using the formulas. Specifically, if the degeneracy $d$ is larger
than the rotary fold $R$ of the symmetry axis: 
\begin{equation}
d>R\text{,}  \label{degeneracy}
\end{equation}
then we assume that the degeneracy will be divided into $\gamma $
sub-degeneracies, where 
\begin{equation}
\gamma =d/R\text{.}  \label{subdegen}
\end{equation}

\item  Strange number $S$: the Strange number $S$ is determined by the
rotary fold $R$ of the symmetry axis \cite{eband} with 
\begin{equation}
S=R-4,  \label{strange}
\end{equation}

where the number $4$ is the highest possible rotary fold number. From Eq. (%
\ref{strange}) and Fig. 1, we get 
\begin{equation}
\begin{tabular}{l}
$\Delta (\Gamma -H)\text{ is a }4\text{-fold rotation axis, }R=4\rightarrow
S=0;$ \\ 
$\Lambda (\Gamma -P)\text{ is a }3\text{-fold rotation axis, }R=3\rightarrow
S=-1;$ \\ 
$\Sigma (\Gamma -N)\text{ is a }2\text{-fold rotation axis, }R=2\rightarrow
S=-2.$%
\end{tabular}
\label{DLS-Strang}
\end{equation}
For the other three symmetry axes $D(P-N)$, $F(P-H)$, and $G(M-N),$ which
are on the surface of the first Brillouin zone (see Fig. 1), we determine
the strange numbers as follows: 
\begin{equation}
\begin{tabular}{l}
$D(P-N)\text{ is parallel to axis }\Delta \text{, }S_{D}=S_{\Delta }=0;$ \\ 
$F\text{ is parallel to an axis equivalent to }\Lambda \text{, }%
S_{F}=S_{\Lambda }=-1;$ \\ 
$G\text{ is parallel to an axis equivalent to }\Sigma \text{, }%
S_{G}=S_{\Sigma }=-2\text{.}$%
\end{tabular}
\end{equation}

\item  Electric charge $Q$: after obtaining $B,$ $S$ and $I$, we can find
the charge $Q$ from the Gell-Mann-Nishijiman relationship \cite{GellMann}: 
\begin{equation}
Q=I_{z}+1/2(S+B).  \label{charge}
\end{equation}

\item  Charmed number $C$ and Bottom number $b$: Since the Lee Particles do
not have any partial charge and the unstable baryons are the energy band
excited states of the Lee Particles (see \textbf{Hypothesis III}), the
unstable baryons shall not have partial charges. Thus, \textbf{if a partial
charge is resulted from (\ref{strange}) and (\ref{charge})}, we have to
consider fluctuation (see \textbf{Hypothesis IV}). The formula (\ref{strange}%
) shall be changed into 
\begin{equation}
\bar{S}=R-4.  \label{strangebar}
\end{equation}

From \textbf{Hypothesis IV (}$\Delta S=\pm 1$\textbf{),} the real value of $%
S $ is 
\begin{equation}
S=\bar{S}+\Delta S=(R-4)\pm 1.  \label{strangeflu}
\end{equation}

The ``Strange number'' $S$ in (\ref{strangeflu}) is not completely the same
as the strange number in (\ref{strange}). In order to compare it with the
experimental results, we would like to give it a new name under certain
circumstances. Based on \textbf{Hypothesis IV}, the new names will be the 
\textbf{Charmed} number and the \textbf{Bottom} number: 
\begin{gather}
\text{if }S=+1\text{ which originates from the fluctuation }\Delta S=+1\text{%
, }  \notag \\
\text{then we call it the \textbf{Charmed} number }C\text{ }(C=+1)\text{;}
\label{charmed}
\end{gather}
\begin{gather}
\text{if }S=-1\text{ which originates from the fluctuation }\Delta S=+1\text{%
, }  \notag \\
\text{and if there is an energy fluctuation,}  \notag \\
\text{then we call it the \textbf{Bottom} number }b\text{ }(b=-1)\text{.}
\label{bottom}
\end{gather}

Thus, (\ref{charge}) needs to be generalized to 
\begin{equation}
Q=I_{z}+1/2(B+S_{G})=I_{z}+1/2(B+S+C+b),  \label{chargeflu}
\end{equation}

where we define the generalized strange number as 
\begin{equation}
S_{G}=S+C+b.  \label{strangegen}
\end{equation}

\item  Charmed strange baryon $\Xi _{C}$ and $\Omega _{C}$: if the energy
band degeneracy $d$ is larger than the rotary fold $R$, the degeneracy will
be divided. Sometimes degeneracies should be divided more than once. After
the first division, the sub-degeneracy energy bands have $S_{Sub}=\bar{S}%
+\Delta S$. For the second division of a degeneracy bands, we have: 
\begin{gather}
\text{if the second division has fluctuation }\Delta S=+1\text{, \ }  \notag
\\
\text{ then }S_{Sub}\text{ may be unchanged and we may have }  \notag \\
\text{ a Charmed number }C\text{ from }C=\Delta S=+1.  \label{C&S}
\end{gather}
Therefore, we can obtain charmed strange baryons $\Xi _{C}$ and $\Omega _{C}$%
.

\item  We assume that a baryon's static mass is the minimum energy of the
energy curved

surface which represents the baryon.
\end{enumerate}

\section{The Energy Bands}

Since the Lee Particle is a Fermion, its motion equation should be the Dirac
equation. Taking into account that (according to the renormalization theory 
\cite{renormal}) the bare mass of the Lee Particle shall be infinite (much
larger than the empirical values of the baryon masses), we use the
Schr\"{o}dinger equation instead of the Dirac equation (our results will
show that this is a very good approximation): 
\begin{equation}
\frac{\text{%
h\hskip-.2em\llap{\protect\rule[1.1ex]{.325em}{.1ex}}\hskip.2em%
}^{2}}{2m_{b}}\nabla ^{2}\Psi +(\varepsilon -V(\vec{r}))\Psi =0,
\label{Schrodinger}
\end{equation}
where $V(\vec{r})$ denotes the strong interaction periodic field\ with body
center cubic symmetries and $m_{b}$ is the bare mass of the Lee Particle.

Using the energy band theory \cite{eband} and the free particle
approximation \cite{freeparticle} (taking $V(\vec{r})=V_{0}$ constant and
making the wave functions satisfy the body center cubic periodic
symmetries), we have \cite{NetXu(E-Band)} 
\begin{equation}
\frac{\text{%
h\hskip-.2em\llap{\protect\rule[1.1ex]{.325em}{.1ex}}\hskip.2em%
}^{2}}{2m_{b}}\nabla ^{2}\Psi +(\varepsilon -V_{0})\Psi =0,  \label{motion}
\end{equation}
where $V_{0}$ is a constant potential. The solution of Eq.(\ref{motion}) is
a plane wave 
\begin{equation}
\Psi _{\vec{k}}(\vec{r})=\exp \{-i(2\pi /a_{x})[(n_{1}-\xi )x+(n_{2}-\eta
)y+(n_{3}-\zeta )z]\},  \label{wave}
\end{equation}
where the wave vector $\vec{k}=(2\pi /a_{x})(\xi ,\eta ,\zeta )$, $a_{x}$ is
the periodic constant, and $n_{1}$, $n_{2}$, $n_{3}$ are integers satisfying
the condition (the result of the periodic symmetries of the body center
cubic field) 
\begin{equation}
n_{1}+n_{2}+n_{3}=\pm \text{ even number or }0.  \label{condition}
\end{equation}
Condition (\ref{condition}) implies that the vector $\vec{n}%
=(n_{1},n_{2},n_{3})$ can only take certain values. For example, $\vec{n}$
can not take $(0,0,1)$ or $\left( 1,1,-1\right) $, but can take $(0,0,2)$
and $(1,-1,2)$.

The zeroth-order approximation of the energy \cite{freeparticle} is 
\begin{equation}
\varepsilon ^{(0)}(\vec{k},\vec{n})=V_{0}+\alpha E(\vec{k},\vec{n}),
\label{mass}
\end{equation}
\begin{equation}
\alpha =h^{2}/2m_{b}a_{x}^{2},  \label{C-ALPHAR}
\end{equation}
\begin{equation}
E(\vec{k},\vec{n})=(n_{1}-\xi )^{2}+(n_{2}-\eta )^{2}+(n_{3}-\zeta )^{2}.
\label{energy}
\end{equation}

Considering the symmetries of the body center cubic periodic field, the wave
functions will satisfy the symmetries of the point group and space group of
the BCC lattice, and the parabolic energy curve of the free Lee Particle
will be changed to energy bands. The wave functions are not needed for the
zeroth order approximation, so we only show the energy bands in Fig. 2-5.
There are six small figures in Fig. 2-4. Each of them shows the energy bands
in one of the six axes in Fig. 1. Each small figure is a schematic one where
the straight lines that show the energy bands shold be parabolic curves. The
numbers above the lines are the values of $\vec{n}$ = ($n_{1}$, $n_{2}$, $%
n_{3}$). The numbers under the lines are the fold numbers of the energy
bands with the same energy (the zeroth order approximation). The numbers
beside both ends of an energy band (the intersection of the energy band line
and the vertical lines) represent the highest and lowest E($\vec{k}$,$\vec{n}
$) values (see Eq. (\ref{energy})) of the band. Putting the values of the E($%
\vec{k}$,$\vec{n}$) into Eq. (\ref{mass}), we get the zeroth order energy
approximation values (in Mev).

\section{The Recognition of the Baryons}

According to \textbf{Hypothesis I}, the nucleons are the ground bands.
Therefore, we can determine $V_{0}$ in formula (\ref{mass})$,$ using the
static masses (static energy) $M_{nucleon}$ of the nucleons. The static
energy ($M_{nucleon}=939$ Mev \cite{particle}) of the nucleons should be the
lowest energy ($V_{0})$ of the energy bands in (\ref{mass}). Thus, at the
ground states, we have 
\begin{equation}
\varepsilon ^{(0)}=V_{0}=M_{nucleon}=939\text{ Mev.}  \label{vzero}
\end{equation}
Fitting the theoretical mass spectrum to the empirical mass spectrum of the
baryons, we can also determine 
\begin{equation}
\alpha =h^{2}/2m_{b}a_{x}^{2}=360\text{ Mev}  \label{alpha}
\end{equation}
in (\ref{mass}). Thus, we have 
\begin{equation}
\varepsilon ^{(0)}(\vec{k},\vec{n})=V_{0}+\alpha E(\vec{k},\vec{n})=939+360E(%
\vec{k},\vec{n})\text{ \ \ (Mev).}  \label{massconst}
\end{equation}
Using \textbf{Hypothesis V} and the energy bands (Fig. 2-5), we can find the
quantum numbers and masses of all excited energy bands. Then, from the
quantum numbers and the masses we can recognize the\textbf{\ unstable
baryons }\cite{NetXu(Baryons)}. As an example, we recognize the \textbf{%
unstable baryons }on the axis $\Delta (\Gamma -H)$.

The axis $\Delta (\Gamma -H)$ is a $4$ fold rotary symmetry axis, $R=4$.
From (\ref{DLS-Strang}), we get the strange number $S=0$. For low energy
levels, there are $8$ and $4$ fold degenerate energy bands and single bands
on the axis. Since the axis has $R=4$, from (\ref{degeneracy}) and (\ref
{subdegen}), the energy bands of degeneracy $8$ will be divided into two 4
fold degenerate bands.

For \textbf{the 4 fold degenerate bands} (see Fig. 2(a) and Fig. 5(a)),
using (\ref{isomax}), we get the isospin $I_{m}=3/2$, and using (\ref{charge}%
), we have $Q=2$, $1$, $0$, $-1$. Comparing them with the experimental
results \cite{particle} that the baryon families $\Delta (\Delta
^{++},\Delta ^{+},\Delta ^{0},\Delta ^{-})$ have $S=0$, $I=3/2$, $Q=2$, $1$, 
$0$, $-1$, we discover that each four fold degenerate band represents a
baryon family $\Delta $. Using (\ref{isonext}), we get\ another $I=3/2-1=1/2$%
, and from (\ref{charge}), we get $Q=1$, $0$. From the facts \cite{particle}
that the baryon families $N(N^{+},N^{0})$ have $S=0$, $I=1/2$, and $Q=1$, $0$%
, we know that there is another baryon family $N$ corresponding to each $%
\Delta $ family. Using Fig. 2(a) and Fig. 5(a), we can get $E_{\Gamma },$ $%
E_{H},$ and $\vec{n}$ values. Then, putting the values of the $E_{\Gamma }$
and $E_{H}$ into the energy formula (\ref{massconst}), we can find the
values of the energy $\varepsilon ^{(0)}$. Finally, we have 
\begin{equation}
\begin{array}{lllll}
E_{H}=1 & \vec{n}=(\text{101,-101,011,0-11}) & \varepsilon ^{(0)}=1299 & 
\Delta (1299); & N(1299) \\ 
E_{\Gamma }=2\text{ } & \vec{n}=(\text{110,1-10,-110,-1-10}) & \varepsilon
^{(0)}=1659 & \Delta (1659); & N(1659) \\ 
E_{\Gamma }=2 & \vec{n}=(\text{10-1,-10-1,01-1,0-1-1}) & \varepsilon
^{(0)}=1659 & \Delta (1659); & N(1659) \\ 
E_{H}=3 & \vec{n}=(\text{112,1-12,-112,-1-12}) & \varepsilon ^{(0)}=2019 & 
\Delta (2019); & N(2019) \\ 
E_{\Gamma }=4 & \vec{n}=(\text{200,-200,020,0-20}) & \varepsilon ^{(0)}=2379
& \Delta (2379); & N(2379) \\ 
E_{H}=5 & \vec{n}=(\text{121,1-21,-121,--1-21}, & \varepsilon ^{(0)}=2739 & 
\Delta (2739); & N(2739) \\ 
& \text{ \ \ \ \ \ \ \ 211,2-11,-211,-2-11}) & \varepsilon ^{(0)}=2739 & 
\Delta (2739); & N(2739) \\ 
E_{H}=5 & \vec{n}=(\text{202,-202,022,0-22}) & \varepsilon ^{(0)}=2739 & 
\Delta (2739); & N(2739) \\ 
E_{H}=5 & \vec{n}=(\text{013,0-13,103,-103}) & \varepsilon ^{(0)}=2739 & 
\Delta (2739); & N(2739) \\ 
\ldots &  &  &  & 
\end{array}
\label{DELTA_4}
\end{equation}

From Fig. 2(a) and Fig. 5(b), we can see that there exist\textbf{\ single
bands }on the axis $\Delta $. From (\ref{isomax}), we have $I=0$. Using (\ref
{strange}) and (\ref{charge}), we get $S=0$ and $Q=0+1/2(S+B)=1/2$ (a
partial charge). According to \textbf{Hypothesis V. 6}, we should use (\ref
{strangeflu}) instead of (\ref{strange}). Therefore, we have 
\begin{equation}
S_{\text{Single}}=\bar{S}_{\Delta }\pm \Delta S=0\pm 1\text{,}
\label{single}
\end{equation}
where $\Delta S=\pm 1$ from \textbf{Hypothesis IV}. The best way to
guarantee the validity of Eq.(\ref{strangebar}) in any small region is to
assume that $\Delta S$ takes $+1$ and $-1$ alternately from the lowest
energy band to higher ones. In fact, the $\vec{n}$ values are really
alternately taking positive and negative values. Using the fact, we can find
a phenomenological formula. If we define a function $Sign(\vec{n})$

\begin{equation}
Sign(\vec{n})=\frac{n_{1}+n_{2}+n_{3}}{\left| n_{1}\right| +\left|
n_{2}\right| +\left| n_{3}\right| }\text{ ,}  \label{Sign}
\end{equation}
then the \textbf{phenomenological formula is}

\begin{equation}
\Delta S=-(1+S_{axis})Sign(\vec{n}).  \label{Dalta-S}
\end{equation}

Before recognizing the baryons, we need to discuss the fluctuation of energy.

The fluctuation of the strange number will be accompanied by an energy
change (\textbf{Hypothesis IV}). We assume that the change of the energy is
proportional to $(\Delta S),$ a number $K\equiv 4-R$ ($R$ is the rotary
number of the axis)$,$ and a number $J$ representing the energy level with a 
\textbf{phenomenological formula:} 
\begin{equation}
\Delta \varepsilon =\left\{ 
\begin{tabular}{l}
$(-1)^{K}100[(J-1)\times K-\delta (K)]\Delta S\text{ \ \ }J=1\text{, }2\text{%
, ...}$ \\ 
$\text{ \ \ \ \ \ \ \ \ }0\text{ \ \ \ \ \ \ \ \ \ \ \ \ \ \ \ \ \ \ \ \ \ \
\ \ \ \ \ \ \ \ \ \ \ \ \ \ \ \ \ \ \ \ \ }J=0\text{\ ,}$%
\end{tabular}
\right\}  \label{eformula}
\end{equation}
where $\delta (K)$ is a Dirac function ($\delta (K)=1$ when $K=0,$ and $%
\delta (K)=0$ when $K\neq 0$.), and $J$ is the energy level number ($J=0,1$, 
$2$, $3,$...) with asymmetric $\vec{n}$ values (or with partial electric
charge from (\ref{strange}) for single energy bands) {}

Due to the fluctuation, the energy formula (\ref{massconst}) should be
changed to 
\begin{eqnarray}
\mathbf{\varepsilon } &=&\mathbf{\varepsilon }^{(0)}\mathbf{(\vec{k},\vec{n}%
)+\Delta \varepsilon }  \notag \\
&=&\mathbf{939+360E(\vec{k},\vec{n})+\Delta \varepsilon }\text{ \ .}
\label{UNITEDMASS}
\end{eqnarray}
\textbf{The formula (\ref{UNITEDMASS}) is the united mass formula which can
give the masses of all the baryons.} It is worth while to emphasize \textbf{%
that the fluctuation of the energy is very small. For the most part of the
baryons, the energy fluctuation is zero (such as (\ref{isomax})). \ The
energy fluctuation is about 5 percent for single energy bands of the axis }$%
\Delta $\textbf{\ and the axis }$\Sigma $\textbf{.\qquad \qquad \qquad
\qquad \qquad \qquad \qquad \qquad \qquad \qquad \qquad \qquad \qquad \qquad
\qquad \qquad }\qquad \qquad \qquad \qquad \qquad \qquad \qquad \qquad
\qquad \qquad \qquad \qquad \qquad \qquad \qquad \qquad \qquad \qquad \qquad
\qquad \qquad \qquad \qquad \qquad \qquad \qquad \qquad \qquad \qquad \qquad
\qquad \qquad \qquad \qquad \qquad \qquad \qquad 

After obtaining the energy fluctuation formula, we come back to the study of
the single bands on the axis $\Delta $.

First, at $E_{\Gamma }=0,J=0,$ $\varepsilon =939$ from (\ref{UNITEDMASS}),
the lowest energy band with $\vec{n}=(0,0,0)$ represents the baryon $N(939)$
from\ (\ref{nucleon}).

Then, at $E_{H}=1,$ the second lowest band with $\vec{n}=(0,0,2)$ and $J=1$.
From (\ref{Dalta-S}), $\Delta S=-1$. Thus $S=-1$, and $\Delta \varepsilon
=100$ Mev from (\ref{eformula})$\rightarrow $ the energy $\varepsilon =1399$
Mev from (\ref{UNITEDMASS}), as well as $I=Q=0$ from (\ref{charge}).
Therefore, it represents the baryon $\Lambda (1399)$.

At $E_{\Gamma }=4$, the band with $\vec{n}=(0,0,-2)$, we get $\Delta S=+1$
from (\ref{Dalta-S}). Thus, $\mathbf{S=\bar{S}}_{\Delta }\mathbf{+1=1}$%
\textbf{. }The energy $\varepsilon =2279$ from (\ref{UNITEDMASS}) and (\ref
{eformula}). Here $S=+1$ originates from the fluctuation $\Delta S=+1$ and
there is an energy fluctuation of $\Delta \varepsilon =-100$. From \textbf{%
Hypothesis V. 6 (\ref{charmed})}, we know the energy band has a charmed
number $C=+1$. \textbf{It represents a new baryon with }$I=0$, $C=+1$, and $%
Q=+1$.\ Since it has a charmed number $C=+1$, \textbf{we will call it the
CHARMED baryon }$\Lambda _{C}^{+}(2279)$ \cite{charmed}. It is very
important to pay attention to \textbf{the Charmed baryon }$\Lambda
_{C}^{+}(2279)$\textbf{\ born here, on the single energy band, and from the
fluctuation } $\Delta S=+1$ and $\Delta \varepsilon =-100$ Mev.

Continuing the above procedure, we have 
\begin{equation}
\begin{array}{llllll}
E_{H}=1 & \vec{n}=(\text{002}) & \Delta S=-1 & J=1 & \Delta \varepsilon =+100
& \Lambda (1399) \\ 
E_{\Gamma }=4 & \vec{n}=(\text{00-2}) & \Delta S=+1 & J=2 & \Delta
\varepsilon =-100 & \Lambda _{C}^{+}(2279) \\ 
E_{H}=9 & \vec{n}=(\text{004}) & \Delta S=-1 & J=3 & \Delta \varepsilon =+100
& \Lambda (4279) \\ 
E_{\Gamma }=16 & \vec{n}=(\text{00-4}) & \Delta S=+1 & J=4 & \Delta
\varepsilon =-100 & \Lambda _{C}^{+}(6599) \\ 
E_{H}=25 & \vec{n}=(\text{006}) & \Delta S=-1 & J=5 & \Delta \varepsilon
=+100 & \Lambda (10039) \\ 
E_{\Gamma }=36 & \vec{n}=(\text{00-6}) & \Delta S=+1 & J=6 & \Delta
\varepsilon =-100 & \Lambda _{C}^{+}(13799) \\ 
\ldots &  &  &  &  & 
\end{array}
\label{DELTA-ONE}
\end{equation}

Continuing above procedure, we can find the whole baryon spectrum \cite
{NetXu(Baryons)}. Our results are shown in Tables $1$ though $6$.

\section{Comparing Results}

We compare the theoretical results of the BCC model to the experimental
results \cite{particle} using Tables 1-6 \cite{NetXu(Results)}. In the
comparison, we will use the following laws:

(1) We do not take into account the angular momenta of the experimental
results. We assume that the small differences of the masses in the same
group of baryons originate from their different angular momenta. If we
ignore this effect, their masses should be essentially the same.

(2) We use the baryon name to represent the intrinsic quantum numbers as
shown in the second column of Table 1.

(3) For low energy cases, the baryons from different symmetry axes with the
same S, C, b, Q, I, I$_{Z}$, and $\overrightarrow{n}$ value, as well as in
the same Brillouin zone are regarded as the same baryon. The mass of the
baryon is the lowest value of their masses.

\ The ground states of various kinds of baryons are shown in Table 1. \
These baryons have a relatively long lifetime and are the most important
experimental results of the baryons. From Table 1, we can see that all
theoretical intrinsic quantum numbers ( $I$, $S$, $C$, $b$, and $Q$) are the
same as experimental results. Also the theoretical mass values are in very
good agreement with the experimental values.

From Table 2-6, we can see that the intrinsic quantum numbers of the
theoretical results are the same as the experimental results.\ Also the
theoretical masses of the baryons\ are in very good agreement with the
experimental results.

The theoretical results $N(1209)$ is not found in experiments. We guess that
it is covered up by the experimental baryon $\Delta (1232)$. The reasons are
as follows : (1) they are unflavored baryons with the same S = C = b =0 and
Q (Q$_{N^{+}}=Q_{\Delta ^{+}}$ and Q$_{N^{0}}=Q_{\Delta ^{0}})$; (2) they
have the same $\overrightarrow{n}$ values ($\overrightarrow{n}%
_{N(1209)}=(011,101)$, $\overrightarrow{n}_{\Delta (1299)}=($%
101,-101,011,0-11$)$) and they are both in the second \textbf{Brillouin zone}%
; (3) the experimental width (120 Mev) of $\Delta (1232)$ is very large, and
the baryon $N(1209)$ is fall within the width region of $\Delta (1232)$; (4)
the mass ($1209$ Mev) of $N(1209)$ is essentially the same as the mass ($%
1232 $ Mev) of $\Delta (1232)$. The facts that the experimental value 1232
is much lower than the theoretical value 1299 of $\Delta (1299)$ and the
experimental width (120) is much larger than other baryons (with similar
masses) support the explanation.

In summary, the BCC model explains all baryon experimental intrinsic quantum
numbers and masses. Virtually no experimentally confirmed baryon is not
included in the model.

\section{Predictions and Discussion}

\subsection{Some New Baryons}

According to the BCC model, a series of possible baryons exist. However,
when energy goes higher and higher, on one hand, the theoretical energy
bands (baryons) will become denser and denser; while on the other hand, the
experimental full widthes of the baryons will become wider and wider. This
makes the possible baryons extremely difficult to be separated. Therefore, \
currently it is very difficult\ to discover higher energy baryons predicted
by the BCC model. We believe that many new baryons will be discovered in the
future with the development of more sensitive experimental techniques. The
following new baryons predicted by the model seem to have a better chance to
be discovered in the not too distant future: $\Lambda ^{0}(2559)$, $\Omega
^{-}(3619)$, $\Lambda _{b}^{0}(10159)$, $\Lambda _{C}^{+}(6599)$, $\Xi
_{C}(3169)$, $\Sigma _{C}(2969)$...

\subsection{ Experimental Verification of the BCC Model}

From Fig. 5 (c), we see three ``brother'' baryons: at E$_{N}=1/2,\ 
\overrightarrow{n}=(1,1,0),$ $\Lambda (1119)$; at E$_{N}=9/2,$ $%
\overrightarrow{n}=(2,2,0),$ $\Lambda (2559)$; at E$_{N}=25/2,$ $%
\overrightarrow{n}=(3,3,0),$ $\Lambda _{b}(5639)$. They are born on the same
symmetry axis $\Sigma $ and at the same symmetry point $N$. The three
``brothers'' have the same isospin I = 0, the same electric charge Q = 0,
and the same generalized strange number (see (\ref{strangegen})) S$_{G}$ = S
+ C + b = -1. Among the three ``brothers'', the light one ($\Lambda (1119)$)
and the heavy one ($\Lambda _{b}(5639))$ both have long lifetimes ($\tau =$
2.6$\times $10$^{-10}s$ for $\Lambda (1119)$, $\tau =$ 1.1$\times $10$^{-12}s
$ for $\Lambda _{b}(5639)$), but the middle one ($\Lambda (2559)$) has not
been discovered. Thus, we propose that one search for the long lifetime
baryon $\Lambda (2559)$ (I = 0, S = -1, Q = 0, M = 2559 Mev, and lifetime 2.6%
$\times $10$^{-10}s>\tau >$1.1$\times $10$^{-12}s$ ). The discovery of the
baryon $\Lambda (2559)$ will provide a strong support for the BCC model.

\subsection{Discussions}

1. From (\ref{alpha}), we have $m_{b}a_{x}^{2}=h^{2}/720$ Mev. Although we
do not know the values of $m_{b}$ and $a_{x}$, we find that $m_{b}a_{x}^{2}$
is a constant. According to the renormalization theory \cite{renormal}, the
bare mass of the Lee Particle should be infinite, so that $a_{x}$ will be
zero. Of course, the infinite and the zero are physical concepts in this
case. We understand that the ``infinite'' means $m_{b}$ is huge and the
``zero'' means $a_{x}$ is much smaller than the nuclear radius. ``$m_{b}$ is
huge'' guarantees that we can use the Schr\"{o}dinger equation (\ref
{Schrodinger}) instead of the Dirac equation, and ``$a_{x}$ is much smaller
than the nuclear radius'' makes the structure of the vacuum material very
difficult to be discovered.

2. The BCC model presents not only a baryon spectrum, but also a reasonable
explanation for the experimental fact that all baryons automatically decay
to nucleons ($p$ or $n$) in a very short time ($<10^{-9}$ second). The
reason is very simple: the baryons are energy band excited states of the Lee
Particle, while the nucleons are the ground band states of the Lee
Particles. It is a well known law in physics that the excited states will
decay into the ground state.

3. After nuclear fusion energy was discovered, we understand the sun's
energy. Similarly, the superconductor will help us explain the vacuum
material. The vacuum material is a super ideal superconductor.

\section{Conclusions\ \ \ }

1. Although baryons ($\Delta ,$ $N,$ $\Lambda ,$ $\Sigma ,$ $\Xi ,$ $\Omega
, $ $\Lambda _{C,}$ $\Xi _{C},$ $\Sigma _{C},$ and $\Lambda _{b}...$) are so
different from one another in I, S, C, b, Q, and M, they may be the same
kind of particles (the Lee Particles), which are in different energy band
states. The long life baryons $\Lambda (1116),$ $\Sigma (1193),$ $\Omega
(1672),$ $...$ may be the metastable states.

2. The quantum number S, C, and b may not come from the quarks (s, c, b),
they may be from symmetries of the body center cubic periodic field. Frank
Wilczek pointed out in Reviews of Modern Physics \cite{Wilczek2}: Some
``appropriate symmetry principles and degrees of freedom, in terms of which
the theory should be formulated, have not yet been identified.''\ We believe
that the body center cubic periodic symmetry of the vacuum material may be
``the appropriate symmetry''.

3. There may be only 2 kinds of quarks (u and d), each of them has three
colored members, in the quark family. The super-strong attractive forces
(color) make the colorless Lee Particle (uud and udd) first. Then the Lee
Particles constitute a body center cubic lattice in the vacuum.

4. Due to the existence of the vacuum material, all observable particles are
constantly affected by the vacuum material (the Lee Particle lattice). Thus,
some laws of statistics (such as fluctuation) can not be ignored.

5. \textbf{There is no other model which can deduce the full baryon spectrum
using a united mass formula.}

6. Although the BCC model successfully explain the baryon spectrum, the
baryon spectrum is deduced from 5 phenomenological Hypotheses and 3
phenomenological formulas in the BCC model. Thus, the BCC model is only a
phenomenological model.

\begin{center}
\bigskip \textbf{Acknowledgment}
\end{center}

I would like to express my heartfelt gratitude to Dr. Xin Yu for checking
the calculations of the energy bands and helping write the paper. I
sincerely thank Professor Robert L. Anderson for his valuable advice. I also
acknowledge\textbf{\ }my indebtedness to Professor D. P. Landau for his
help. I specially thank Professor T. D. Lee, for his \textit{particle physics%
} class in Beijing and his CUSPEA program which gave me an opportunity for
getting my Ph.D. I thank Professor W. K. Ge very much for his valuable help
and for recommending Wilczek's paper \cite{wilczek}. I thank my friend Z. Y.
Wu very much for his help in preparing the paper. I thank my classmate J. S.
Xie very much for checking the calculations of the energy bands. I thank
Professor Y. S. Wu, H. Y. Guo, and S. Chen \cite{XUarticle} very much for
many very useful discussions.

\begin{center}
\bigskip {\LARGE FIGURES}
\end{center}

Fig. 1. \ The first Brillouin zone of the body center cubic lattice. The the
symmetry points and axes are indicated. The center of the first Brillouin
zone is at the point $\Gamma $. The axis $\Delta $ is a $4$ fold rotation
axis, the strange number S = 0, the baryon family $\Delta $ ($\Delta ^{++},$ 
$\Delta ^{+},$ $\Delta ^{0},$ $\Delta ^{-}$) and $N$ will appear on the
axis. The axes $\Lambda $ and $F$ are $3$ fold rotation axes, the strange
number S =\ -1, the baryon family $\Sigma $ ($\Sigma ^{+},$ $\Sigma ^{0},$ $%
\Sigma ^{-})$ and $\Lambda $ will appear on the axes. The axes $\Sigma $ and
G are $2$ fold rotation axes, the strange number S = -2, the baryon family $%
\Xi $ ($\Xi ^{0},$ $\Xi ^{-}$) will appear on the axes. The axis D is
parallel to the axis $\Delta $, S = 0. And the axis is a $2$ fold rotation
axis, the baryon family N (N$^{+}$, N$^{0}$) will be on the axis.

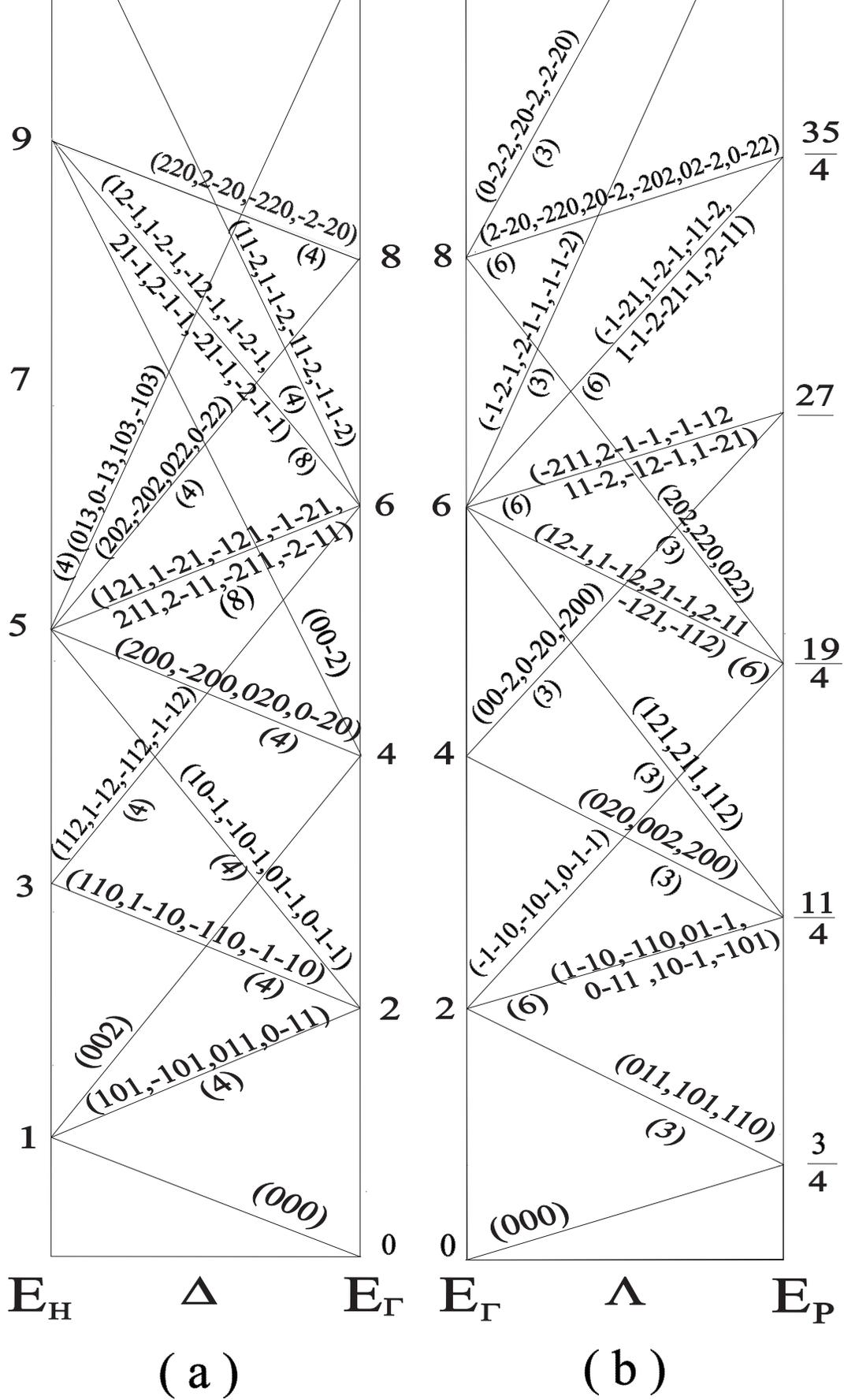
Fig. 2. \ \ (a) The energy bands on the axis $\Delta $ (the axis $\Gamma -$%
H). E$_{\Gamma }$ is the value of E($\vec{k}$, $\vec{n}$) (see Eq. (\ref
{energy})) at the end point $\Gamma ,$ while E$_{H}$ is the value of E($\vec{%
k}$, $\vec{n}$) at other end point H. \ \ (b) The energy bands on the axis $%
\Lambda $ (the axis $\Gamma $-P). E$_{\Gamma }$ is the value of E($\vec{k}$, 
$\vec{n}$) (see Eq. (\ref{energy})) at the end point $\Gamma ,$ while E$_{P}$
is the value of E($\vec{k}$, $\vec{n}$) at other end point P.

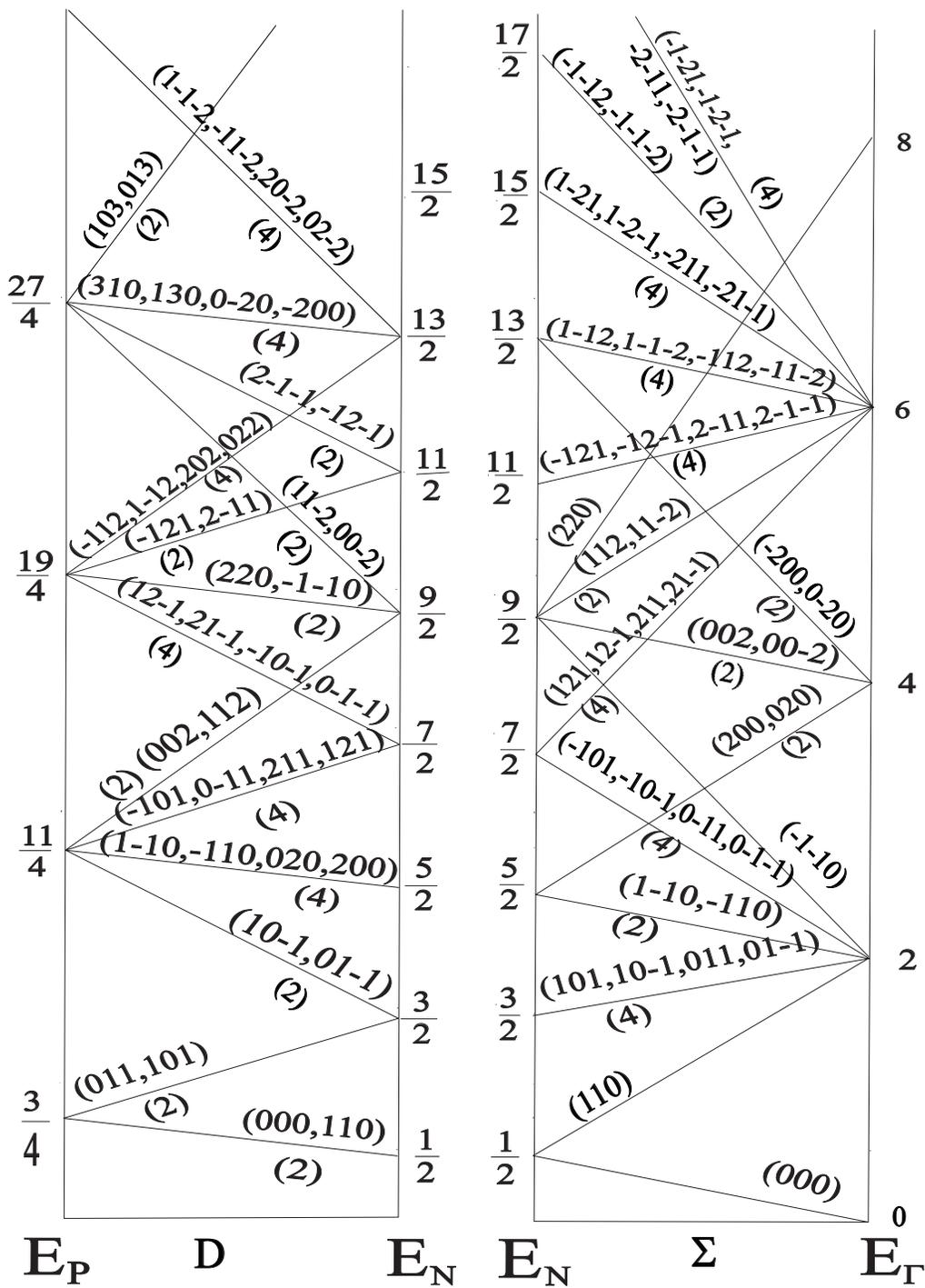
Fig. 3. \ \ (a) The energy bands on the axis $\Sigma $ (the axis $\Gamma $%
-N). E$_{\Gamma }$ is the value of E($\vec{k}$, $\vec{n}$) (see Eq. (\ref
{energy})) at the end point $\Gamma ,$ while E$_{N}$ is the value of E($\vec{%
k}$, $\vec{n}$) at other end point N. \ (b) The energy bands on the axis $D$
(the axis P-N). E$_{P}$ is the value of E($\vec{k}$, $\vec{n}$) (see Eq. (%
\ref{energy})) at the end point P$,$ while E$_{N}$ is the value of E($\vec{k}
$, $\vec{n}$) at other end point N.

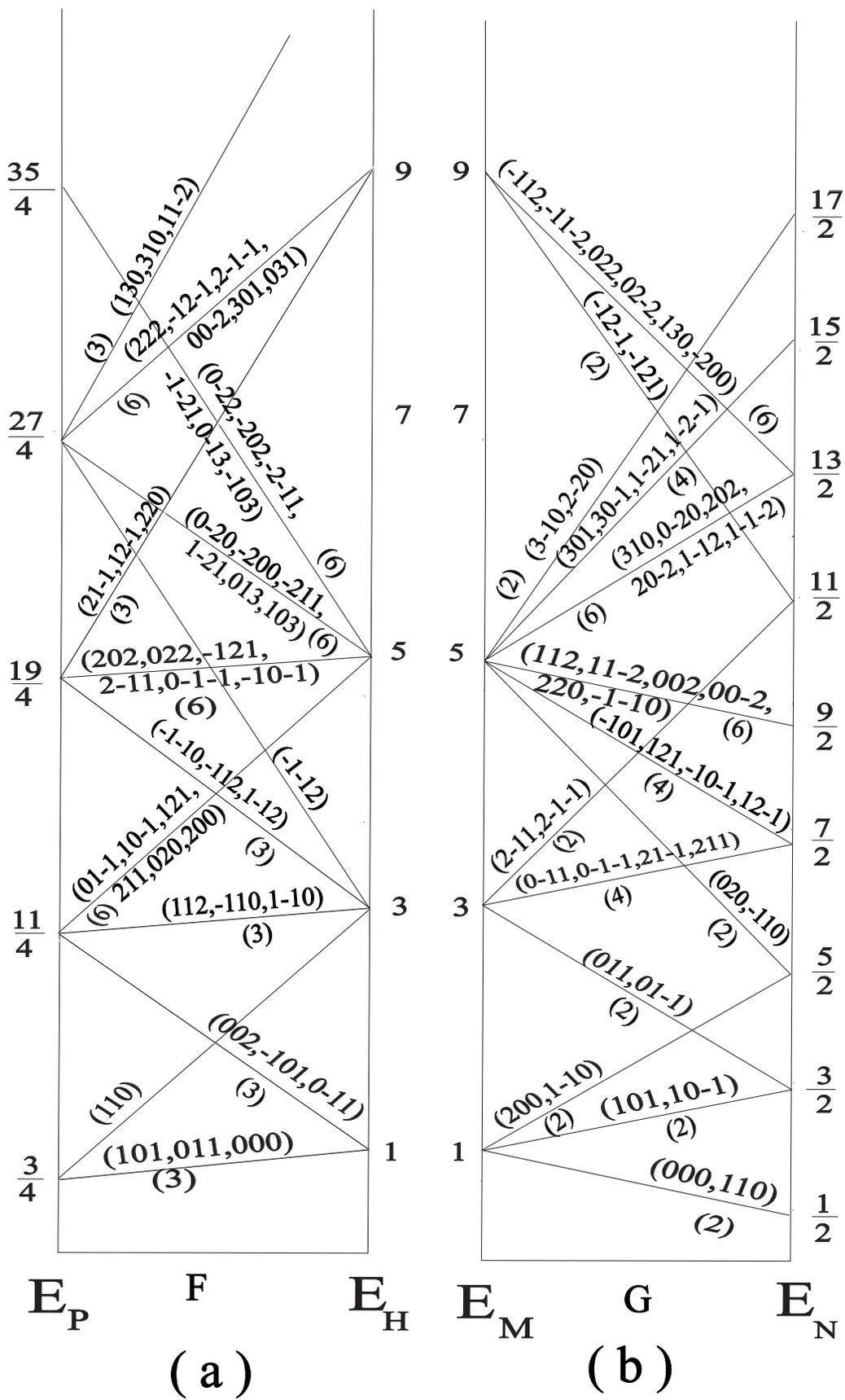
Fig. 4. \ \ (a) The energy bands on the axis $F$ (the axis P-H). E$_{P}$ is
the value of E($\vec{k}$, $\vec{n}$) (see Eq. (\ref{energy})) at the end
point $P,$ while E$_{H}$ is the value of E($\vec{k}$, $\vec{n}$) at other
end point H.\ \ (b) The energy bands on the axis $G$ (the axis M-N).\ E$_{M}$
is the value of E($\vec{k}$, $\vec{n}$) (see Eq. (\ref{energy})) at the end
point M$,$ while E$_{N}$ is the value of E($\vec{k}$, $\vec{n}$) at other
end point N.

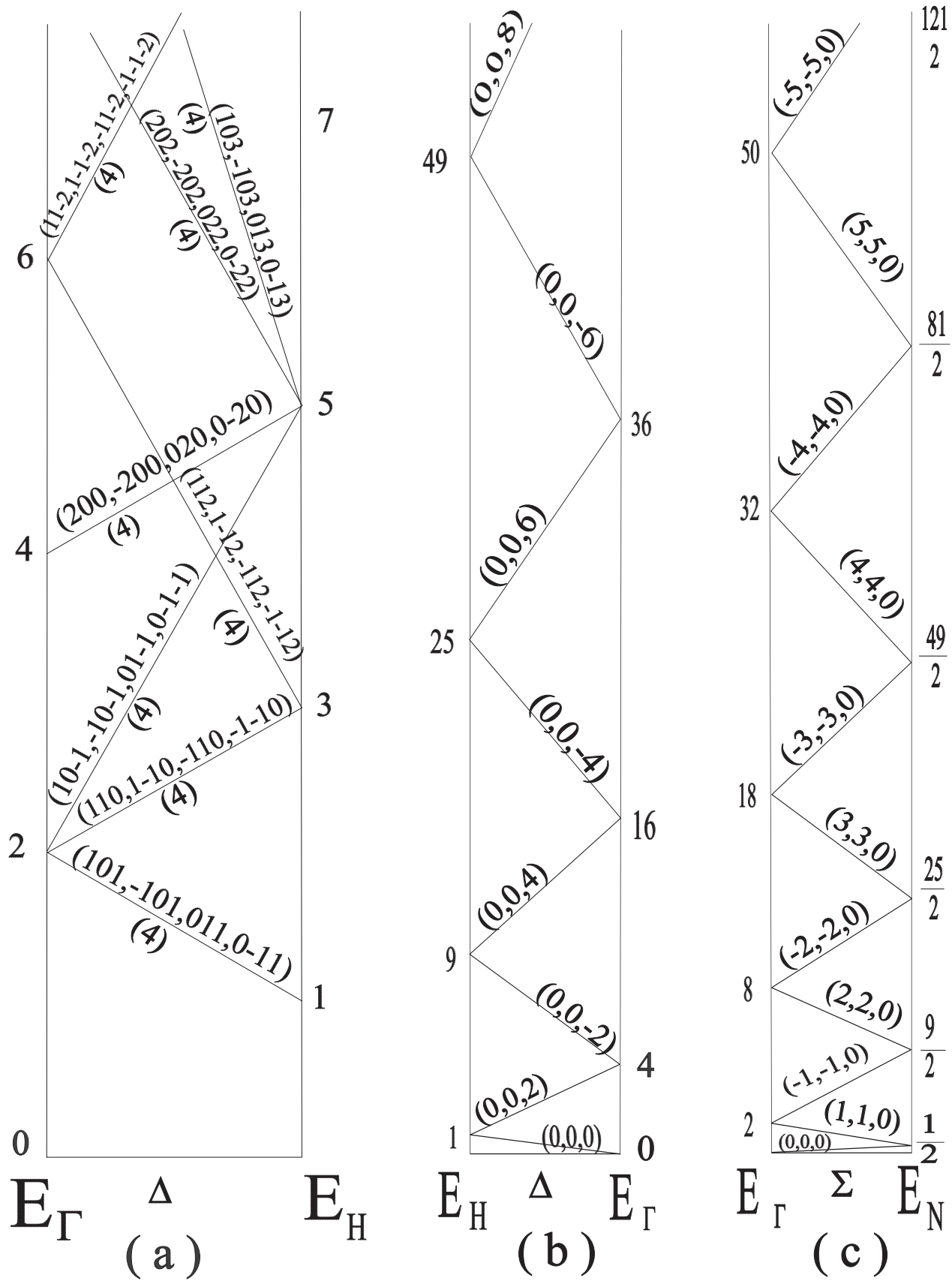
Fig. 5. \ \ (a) The $4$ fold degenerate energy bands (selected from Fig.
2(a)) on the axis $\Delta $ (the axis $\Gamma $-H).\ \ (b) The single energy
bands (selected from Fig. 2(a)) on the axis $\Delta $ (the axis $\Gamma $%
-H). \ \ (c) The single energy band (selected from Fig. 3(a)) on the axis $%
\Sigma $ (the axis $\Gamma $-N).

\newpage \pagebreak

\bigskip

\bigskip

\bigskip

\bigskip

\bigskip

\bigskip

\bigskip

\bigskip

\bigskip

\bigskip

\bigskip

\bigskip

\newpage

\begin{center}
{\LARGE TABLE}
\end{center}

\qquad \qquad \qquad \qquad Table 1. \ The Ground States of the Various
Baryons.

\begin{tabular}{|l|l|l|l|l|}
\hline
{\small Theory} & Quantum. No & {\small Experiment} & R & Life Time \\ \hline
Name({\small M}) & S, C, \ b, \ \ I, \ Q & Name({\small M}) &  &  \\ \hline
p(939) & 0,\ \ 0, \ 0, {\small 1/2, \ \ }1 & p(938) & 0.1 & $ > $ %
10$^{31}years$ \\ \hline
n(939) & 0, \ 0, \ 0, {\small 1/2, \ \ }0 & n(940) & 0.1 & 1.0$\times 10^{8}$
s \\ \hline
$\Lambda (1119)$ & -1, \ 0, \ 0, \ \ 0, \ 0 & $\Lambda (1116)$ & 0.3 & 2.6$%
\times $10$^{-10}s$ \\ \hline
$\Sigma (1209)^{+}$ & -1, \ 0, \ 0, \ \ 1, \ 1 & $\Sigma (1189)^{+}$ & 1.7 & 
.80$\times $10$^{-10}s$ \\ \hline
$\Sigma (1209)^{0}$ & -1, \ 0, \ 0, \ \ 1, \ 0 & $\Sigma (1193)^{0}$ & 1.4 & 
7.4$\times $10$^{-20}s$ \\ \hline
$\Sigma (1209)^{-}$ & -1, \ 0, \ 0, \ \ 1, -1 & $\Sigma (1197)^{-}$ & 1.0 & 
1.5$\times $10$^{-10}s$ \\ \hline
$\Xi (1299)^{0}$ & -2, \ 0, \ 0, {\small 1/2}, 0 & $\Xi (1315)^{0}$ & 1.2 & 
2.9$\times $10$^{-10}s$ \\ \hline
$\Xi (1299)^{-}$ & -2, \ 0, \ 0, {\small 1/2}, -1 & $\Xi (1321)^{-}$ & 1.7 & 
1.6$\times $10$^{-10}s$ \\ \hline
$\Omega (1659)^{-}$ & -3, \ 0, \ 0. \ \ 0, -1 & $\Omega (1672)^{-}$ & 0.8 & 
.82$\times $10$^{-10}s$ \\ \hline
$\Lambda _{c}^{+}(2279)$ & 0, \ 1, \ 0, \ \ 0, \ 1 & $\Lambda _{c}^{+}(2285)$
& 0.3 & .21$\times $10$^{-12}s$ \\ \hline
$\Xi _{c}^{+}(2549)$ & -1, \ 1, \ 0, {\small 1/2}, 1 & $\Xi _{c}^{+}(2466)$
& 3.4 & .35$\times $10$^{-12}$ \\ \hline
$\Xi _{c}^{0}(2549)$ & -1, \ 1, \ 0, {\small 1/2}, 1 & $\Xi _{c}^{0}(2470)$
& 3.2 & .10$\times $10$^{-12}s$ \\ \hline
$\Sigma _{c}^{++}(2449)$ & 0, \ 1, \ 0, \ 1, \ \ 2 & $\Sigma _{c}^{++}(2453)$
& 0.2 &  \\ \hline
$\Sigma _{c}^{+}(2449)$ & 0, \ 1, \ 0, \ 1, \ \ 1 & $\Sigma _{c}^{+}(2454)$
& 0.2 &  \\ \hline
$\Sigma _{c}^{0}(2449)$ & 0, \ 1, \ 0, \ 1, \ \ 0 & $\Sigma _{c}^{0}(2452)$
& 0.1 &  \\ \hline
$\Omega _{c}(2759)$ & 0, \ 0, -1, \ 0, \ 0 & $\Omega _{c}(2704)$ & 2.0 & .64$%
\times $10$^{-13}s$ \\ \hline
$\Lambda _{b}(5639)$ & 0, \ 0, -1, \ 0, \ 0 & $\Lambda _{b}(5641)$ & .04 & 
1.1$\times $10$^{-12}s$ \\ \hline
$\Delta (1299)^{++}$ & 0, \ 0, \ 0, {\small 3/2}, \ 2 & $\Delta (1232)^{++}$
& 5.2 & $\Gamma $=120 Mev \\ \hline
$\Delta (1299)^{+}$ & 0, \ 0, \ 0, {\small 3/2}, \ 1 & $\Delta (1232)^{+}$ & 
5.4 & $\Gamma $=120 Mev \\ \hline
$\Delta (1299)^{0}$ & 0, \ 0, \ 0, {\small 3/2}, \ 0 & $\Delta (1232)^{0}$ & 
5.4 & $\Gamma $=120 Mev \\ \hline
$\Delta (1299)^{-}$ & 0, \ 0, \ 0, {\small 3/2}, -1 & $\Delta (1232)^{-}$ & 
5.4 & $\Gamma $=120 Mev \\ \hline
\end{tabular}

In the fourth column, R =($\frac{\Delta \text{M}}{\text{M}}$){\small \%.}

\bigskip

\bigskip

\bigskip

\bigskip

\bigskip

\bigskip

\bigskip

\bigskip

\bigskip

\bigskip

\bigskip

\bigskip

\bigskip

\bigskip

\bigskip

\bigskip \newpage \qquad\ \ \ \ \ \ \ \ \ \ Table 2. The Unflavored Baryons $%
N$ and $\Delta $ ($S$= $C$=$b$ = 0) \ 

\begin{tabular}{|l|l|l||l|l|l|}
\hline
Theory & Experiment & $\frac{\Delta \text{M}}{\text{M}}\%$ & Theory & 
Experiment & $\frac{\Delta \text{M}}{\text{M}}\%$ \\ \hline
$\mathbf{\bar{N}(939)}$ & $\mathbf{\bar{N}(939)}$ & \textbf{0.0} & $\mathbf{%
\bar{\Delta}(1254)}^{\#}$ & $\mathbf{\bar{\Delta}(1232)}$ & \textbf{1.8} \\ 
\hline
$N(1479)$ & 
\begin{tabular}{l}
$N(1440)$ \\ 
$N(1520)$ \\ 
$N(1535)$%
\end{tabular}
&  &  &  &  \\ \hline
$\mathbf{\bar{N}(1479)}$ & $\mathbf{\bar{N}(1498)}$ & \textbf{1.2} &  &  & 
\\ \hline
\begin{tabular}{l}
$N(1659)$ \\ 
$N(1659)$%
\end{tabular}
& 
\begin{tabular}{l}
$N(1650)$ \\ 
$N(1675)$ \\ 
$N(1680)$ \\ 
$N(1700)$ \\ 
$N(1710)$ \\ 
$N(1720)$%
\end{tabular}
&  & 
\begin{tabular}{l}
$\Delta (1659)$ \\ 
$\Delta (1659)$%
\end{tabular}
& 
\begin{tabular}{l}
$\Delta (1600)$ \\ 
$\Delta (1620)$ \\ 
$\Delta (1700)$%
\end{tabular}
&  \\ \hline
$\mathbf{\bar{N}(1659)}$ & $\mathbf{\bar{N}(1689)\ \ }$ & \textbf{1.7} & $%
\mathbf{\bar{\Delta}(1659)}$ & $\mathbf{\bar{\Delta}(1640)}$ & \textbf{1.2}
\\ \hline
\begin{tabular}{l}
$N(1839)$ \\ 
$N(1839)$ \\ 
$N(1929)$ \\ 
$N(1929)$ \\ 
$N(2019)$%
\end{tabular}
& 
\begin{tabular}{l}
$N(1900)\ast $ \\ 
$N(1990)\ast $ \\ 
$N(2000)\ast $ \\ 
$N(2080)\ast $%
\end{tabular}
&  & 
\begin{tabular}{l}
$\Delta (1929)$ \\ 
$\Delta (1929)$ \\ 
$\Delta (2019)$%
\end{tabular}
& 
\begin{tabular}{l}
$\Delta (1900)$ \\ 
$\Delta (1905)$ \\ 
$\Delta (1910)$ \\ 
$\Delta (1920)$ \\ 
$\Delta (1930)$ \\ 
$\Delta (1950)$%
\end{tabular}
&  \\ \hline
$\mathbf{\bar{N}(1914)}$ & $\mathbf{\bar{N}(1923)}$ & \textbf{0.5} & $%
\mathbf{\bar{\Delta}(1959)}$ & $\mathbf{\bar{\Delta}(1919)}$ & \textbf{2.1}
\\ \hline
\begin{tabular}{l}
$N(2199)$ \\ 
$N(2199)$%
\end{tabular}
& 
\begin{tabular}{l}
$N(2190)$ \\ 
$N(2220)$ \\ 
$N(2250)$%
\end{tabular}
&  &  &  &  \\ \hline
$\mathbf{\bar{N}(2199)}$ & $\mathbf{\bar{N}(2220)}$ & \textbf{0.9} &  &  & 
\\ \hline
\begin{tabular}{l}
$N(2379)$ \\ 
$N(2549)$%
\end{tabular}
&  &  & $\Delta (2379)$ & $\Delta (2420)$ &  \\ \hline
$\ 
\begin{tabular}{l}
$N(2549)$ \\ 
$N(2559)$%
\end{tabular}
$ &  &  & $\mathbf{\bar{\Delta}(2379)}$ & $\mathbf{\bar{\Delta}(2420)}$ & 
\textbf{1.6} \\ \hline
$\mathbf{3N(2649)}$ & $\mathbf{N(2600)}$ & \textbf{1.9} & $\Delta (2649)$ & 
&  \\ \hline
4$N(2739)$ &  &  & 5$\Delta (2739)$ &  &  \\ \hline
\end{tabular}

\#{\small The average of N(1209) and }$\Delta (1299).$

*{\small Evidences are fair, they are not listed in the Baryon Summary Table 
\cite{particle}}.

\bigskip

\bigskip

\bigskip

\bigskip

\bigskip

\bigskip

\bigskip

\bigskip

\bigskip

\bigskip

\bigskip

\bigskip

\bigskip

\bigskip

\bigskip

\bigskip

\bigskip

\bigskip \newpage\ \ \ \ \ \qquad\ \ \ Table 3. Two Kinds of Strange Baryons 
$\Lambda $ and $\Sigma $ ($S=-1$)

\begin{tabular}{|l|l|l||l|l|l|}
\hline
Theory & Experiment & $\frac{\Delta \text{M}}{\text{M}}\%$ & Theory & 
Experiment & $\frac{\Delta \text{M}}{\text{M}}\%$ \\ \hline
$\mathbf{\Lambda (1119)}$ & $\mathbf{\Lambda (1116)}$ & \textbf{0.36} & $%
\mathbf{\Sigma (1209)}$ & $\mathbf{\Sigma (1193)}$ & \textbf{1.4} \\ \hline
\begin{tabular}{l}
$\Lambda (1299)$ \\ 
$\Lambda (1399)$%
\end{tabular}
&  &  &  &  &  \\ \hline
$\overline{\mathbf{\Lambda }}\mathbf{(1349)}$ & $\overline{\mathbf{\Lambda }}%
\mathbf{(1405)}$ & \textbf{4.0} & $\mathbf{\Sigma (1299)}$ & $\mathbf{\Sigma
(1385)}$ & \textbf{6.2} \\ \hline
\begin{tabular}{l}
$\Lambda (1659)$ \\ 
$\Lambda (1659)$ \\ 
$\Lambda (1659)$%
\end{tabular}
& 
\begin{tabular}{l}
$\Lambda (1520)$ \\ 
$\Lambda (1600)$ \\ 
$\Lambda (1670)$ \\ 
$\Lambda (1690)$%
\end{tabular}
&  & 
\begin{tabular}{l}
$\Sigma (1659)$ \\ 
$\Sigma (1659)$ \\ 
$\Sigma (1659)$%
\end{tabular}
& 
\begin{tabular}{l}
$\Sigma (1660)$ \\ 
$\Sigma (1670)$ \\ 
$\Sigma (1750)$ \\ 
$\Sigma (1775)$%
\end{tabular}
&  \\ \hline
$\mathbf{\bar{\Lambda}(1659)}$ & $\mathbf{\bar{\Lambda}(1620)}$ & \textbf{2.4%
} & $\mathbf{\bar{\Sigma}(1659)}$ & $\mathbf{\bar{\Sigma}(1714)}$ & \textbf{%
3.2} \\ \hline
\begin{tabular}{l}
$\Lambda (1929)$ \\ 
$\Lambda (1929)$ \\ 
$\Lambda (1929)$%
\end{tabular}
& 
\begin{tabular}{l}
$\Lambda (1800)$ \\ 
$\Lambda (1810)$ \\ 
$\Lambda (1820)$ \\ 
$\Lambda (1830)$ \\ 
$\Lambda (1890)$%
\end{tabular}
&  & 
\begin{tabular}{l}
$\Sigma (1929)$ \\ 
$\Sigma (1929)$%
\end{tabular}
& 
\begin{tabular}{l}
$\Sigma (1915)$ \\ 
$\Sigma (1939)$%
\end{tabular}
&  \\ \hline
$\mathbf{\bar{\Lambda}(1929)}$ & $\mathbf{\bar{\Lambda}(1830)}$ & \textbf{5.4%
} & $\mathbf{\bar{\Sigma}(1929)}$ & $\mathbf{\bar{\Sigma}(1928)}$ & \textbf{%
.05} \\ \hline
\begin{tabular}{l}
$\Lambda (2019)$ \\ 
$\Lambda (2019)$%
\end{tabular}
& 
\begin{tabular}{l}
$\Lambda (2100)$ \\ 
$\Lambda (2110)$%
\end{tabular}
&  & $\mathbf{\Sigma (2019)}$ & $\mathbf{\Sigma (2030)}$ & \textbf{.54} \\ 
\hline
$\mathbf{\bar{\Lambda}(2019)}$ & $\mathbf{\Lambda (2105)}$ & \textbf{4.1} & 
&  &  \\ \hline
$
\begin{tabular}{l}
$\Lambda (2359)$ \\ 
$\Lambda (2379)$%
\end{tabular}
$ & $\Lambda (2350)$ &  & $\Sigma (2379)$ & 
\begin{tabular}{l}
$\Sigma (2250)$ \\ 
$\Sigma (2455)^{\ast }$%
\end{tabular}
&  \\ \hline
$\mathbf{\bar{\Lambda}(2369)}$ & $\mathbf{\bar{\Lambda}(2350)}$ & \textbf{0.8%
} & $\mathbf{\bar{\Sigma}(2379)}$ & $\mathbf{\bar{\Sigma}(2353)}$ & \textbf{%
1.1} \\ \hline
$\mathbf{\Lambda (2559)}$ & $\mathbf{\Lambda (2585)}^{\ast }$ & \textbf{0.9}
&  &  &  \\ \hline
5$\Lambda (2649)$ &  &  & \textbf{3} $\mathbf{\Sigma (2649)}$ & $\mathbf{%
\Sigma (2620)}$ & \textbf{1.1} \\ \hline
&  &  & 4 $\mathbf{\Sigma (2739)}$ &  &  \\ \hline
\end{tabular}
\ \ \ \ \ \ \ \ \ \ \ \ \ \ \ \ \ \ \ \ \ \ \ \ \ \ \ \ \ \ \ 

{\small \ }*Evidences of existence for these baryons are only fair, they are
not

listed in the Baryon Summary Table \cite{particle}.

\bigskip

\bigskip \newpage

\qquad\ \ \ \ \ \ \ Table 4. The Baryons $\Xi $ and the Baryons $\Omega $

\begin{tabular}{|l|l|l||l|l|l|}
\hline
Theory & Experiment & $\frac{\Delta \text{M}}{\text{M}}\%$ & Theory & 
Experiment & $\frac{\Delta \text{M}}{\text{M}}\%$ \\ \hline
$\ \mathbf{\Xi (1299)}$ & $\mathbf{\Xi (1318)}$ & \textbf{1.5} & $\mathbf{%
\Omega (1659)}$ & $\mathbf{\Omega (1672)}$ & 0\textbf{.8} \\ \hline
$\ \mathbf{\Xi (1479)}$ & $\mathbf{\Xi (1530)}$ & \textbf{3.3} & $\Omega
(2359)$ & 
\begin{tabular}{l}
$\Omega (2250)$ \\ 
$\Omega (2380)$ \\ 
$\Omega (2470)$%
\end{tabular}
&  \\ \hline
\textbf{3}$\mathbf{\Xi (1659)}$ & $\mathbf{\Xi (1690)}$ & \textbf{1.8} & $%
\mathbf{\bar{\Omega}(2359)}$ & $\mathbf{\bar{\Omega}(2367)}$ & \textbf{0.4}
\\ \hline
$\ \mathbf{\Xi (1839)}$ & $\mathbf{\Xi (1820)}$ & \textbf{1.1} & $\Omega
(2879)$ &  &  \\ \hline
\textbf{\ }$\mathbf{\Xi (1929)}$ & $\mathbf{\Xi (1950)}$ & \textbf{1.1} & $%
\Omega (3619)$ &  &  \\ \hline
\textbf{2}$\mathbf{\Xi (2019)}$ & $\mathbf{\Xi (2030)}$ & \textbf{1.0} & $%
\Omega (7019)$ &  &  \\ \hline
\textbf{2}$\mathbf{\Xi (2199)}$ & $\mathbf{\Xi (2250)}^{\ast }$ & \textbf{2.3%
} &  &  &  \\ \hline
$\ \ \mathbf{\Xi (2379)}$ & $\mathbf{\Xi (2370)}^{\ast }$ & \textbf{0.4} & 
&  &  \\ \hline
$\ \Xi (2559)$ &  &  &  &  &  \\ \hline
5$\Xi (2739)$ &  &  &  &  &  \\ \hline
\end{tabular}

\bigskip *Evidences of existence for these baryons are only fair, they are
not

listed in the Baryon Summary Table \cite{particle}.

{\small \bigskip }\newpage

\qquad \qquad Table 5. Charmed \ $\Lambda _{c}^{+}$\ and Bottom $\Lambda _{b%
\text{ }}^{0}$ Baryons \ 

\begin{tabular}{|l|l|l||l|l|l|}
\hline
Theory & Experiment & $\frac{\Delta \text{M}}{\text{M}}\%$ & Theory & 
Experiment & $\frac{\Delta \text{M}}{\text{M}}\%$ \\ \hline
$\mathbf{\Lambda }_{c}^{+}\mathbf{(2279)}$ & $\mathbf{\Lambda }_{c}^{+}%
\mathbf{(2285)}$ & \textbf{0.22} & $\mathbf{\Lambda }_{b}^{0}\mathbf{(5639)}$
& $\mathbf{\Lambda }_{b}^{0}\mathbf{(5641)}$ & 0.035 \\ \hline
$
\begin{tabular}{l}
$\Lambda _{c}^{+}(2449)$ \\ 
$\Lambda _{c}^{+}(2539)$%
\end{tabular}
$ & 
\begin{tabular}{l}
$\Lambda _{c}^{+}(2593)$ \\ 
$\Lambda _{c}^{+}(2625)$%
\end{tabular}
&  & $\Lambda _{b}^{0}(10159)$ &  &  \\ \hline
$\mathbf{\bar{\Lambda}}_{c}^{+}\mathbf{(2494)}$ & $\mathbf{\bar{\Lambda}}%
_{c}^{+}\mathbf{(2609)}$ & \textbf{4.4} &  &  &  \\ \hline
$\Lambda _{c}^{+}(2759)$ &  &  &  &  &  \\ \hline
$\Lambda _{c}^{+}(2969)$ &  &  &  &  &  \\ \hline
$\Lambda _{c}^{+}(6599)$ &  &  &  &  &  \\ \hline
\end{tabular}

\bigskip 

\bigskip

\newpage

\qquad \qquad Table 6. Charmed Strange Baryon $\Xi _{c}$, $\Sigma _{c}$ and $%
\Omega _{C}$ \ 

\begin{tabular}{|l|l|l||l|l|l|}
\hline
Theory & Experiment & $\frac{\Delta \text{M}}{\text{M}}\%$ & Theory & 
Experiment & $\frac{\Delta \text{M}}{\text{M}}\%$ \\ \hline
$\Xi _{c}(2549)$ & 
\begin{tabular}{l}
$\Xi _{c}(2468)$ \\ 
$\Xi _{c}(2645)$%
\end{tabular}
&  & 
\begin{tabular}{l}
$\Sigma _{c}(2449)$ \\ 
$\Sigma _{c}(2539)$%
\end{tabular}
& $
\begin{tabular}{l}
$\Sigma _{c}(2455)$ \\ 
$\Sigma _{c}(2530)\ast $%
\end{tabular}
$ &  \\ \hline
$\mathbf{\bar{\Xi}}_{c}\mathbf{(2549)}$ & $\mathbf{\bar{\Xi}}_{c}\mathbf{%
(2557)}$ & \textbf{0.3} & $\mathbf{\bar{\Sigma}}_{c}\mathbf{(2495)}$ & $%
\mathbf{\bar{\Sigma}}_{c}\mathbf{(2493)}$ & \textbf{0.08} \\ \hline
$\Xi _{C}(3169)$ &  &  & $\Sigma _{c}(2969)$ &  &  \\ \hline
&  &  &  &  &  \\ \hline
&  &  & $\mathbf{\Omega }_{C}\mathbf{(2759)}$ & $\mathbf{\Omega }_{C}\mathbf{%
(2704)}$ & \textbf{2.0} \\ \hline
&  &  & $\Omega _{C}(3679)$ &  &  \\ \hline
\end{tabular}
\ \ \ 

*Evidences of existence for these baryons are only fair, they are not

listed in the Baryon Summary Table \cite{particle}.

\bigskip

\bigskip

\bigskip

\bigskip

\end{document}